\documentclass[preprint,3p,onecolumn,pra,aps,showpacs,superscriptaddress,article]{elsarticle}
\usepackage[english]{babel}
\usepackage[utf8]{inputenc}
\usepackage{amsmath,amssymb,amsthm}
\usepackage{xcolor}
\usepackage{float}
\usepackage[export]{adjustbox} 
\usepackage{textcomp}
\usepackage{amssymb}
\usepackage{graphicx}
\usepackage{esint}
\usepackage{lineno,hyperref}
\usepackage[toc,page]{appendix}
\usepackage{tabularx}
\def\<{\left\langle}
\def\>{\right\rangle}
\def\({\left(}
\def\){\right)}

 \journal{Physica A}
\DeclareUnicodeCharacter{2217}{*} 
\begin{document}





\begin{frontmatter}

\title{Bridging the divide: Economic exchange and segregation in dual-income cities}


\author[1]{Diego Ortega}
\ead{dortega144@alumno.uned.es}

\author[1,2]{Elka Korutcheva}

\affiliation[1]{organization={Dto. Física Fundamental, Universidad Nacional de Educación a Distancia},
city={Madrid},
postcode={28040},
country={Spain}}

\affiliation[2]{organization={G. Nadjakov Institute of Solid State Physics, Bulgarian
  Academy of Sciences},
city={Sofía},
postcode={1784},
country={Bulgaria}}


\begin{abstract}
Segregation is a growing concern around the world. One of its main manifestations is the creation of ghettos, whose inhabitants have difficult access to well-paid jobs, which are often located far from their homes. In order to study this phenomenon, we propose an extension of Schelling's model of segregation to take into account the existence of economic exchanges.  To approximate a geographical model of the city, we consider a small-world network with a defined real estate market. The evolution of the system has also been studied, finding that economic exchanges follow exponential laws and relocations are approximated by power laws. In addition to this, we consider the existence of delays in the actions of the agents, which mainly affect the happiness of those with fewer economic resources. Besides, the size of the economic exchange plays a crucial role in overall segregation. Despite its simplicity, we find that our model reproduces real-world situations such as the separation between favoured and handicapped economic areas, the importance of economic contacts between them to improve the distribution of wealth, and the existence of efficient and cheap transport to break the poverty cycles typical of disadvantaged zones.

\end{abstract}

\begin{graphicalabstract}
\center
\includegraphics[width=16 cm]{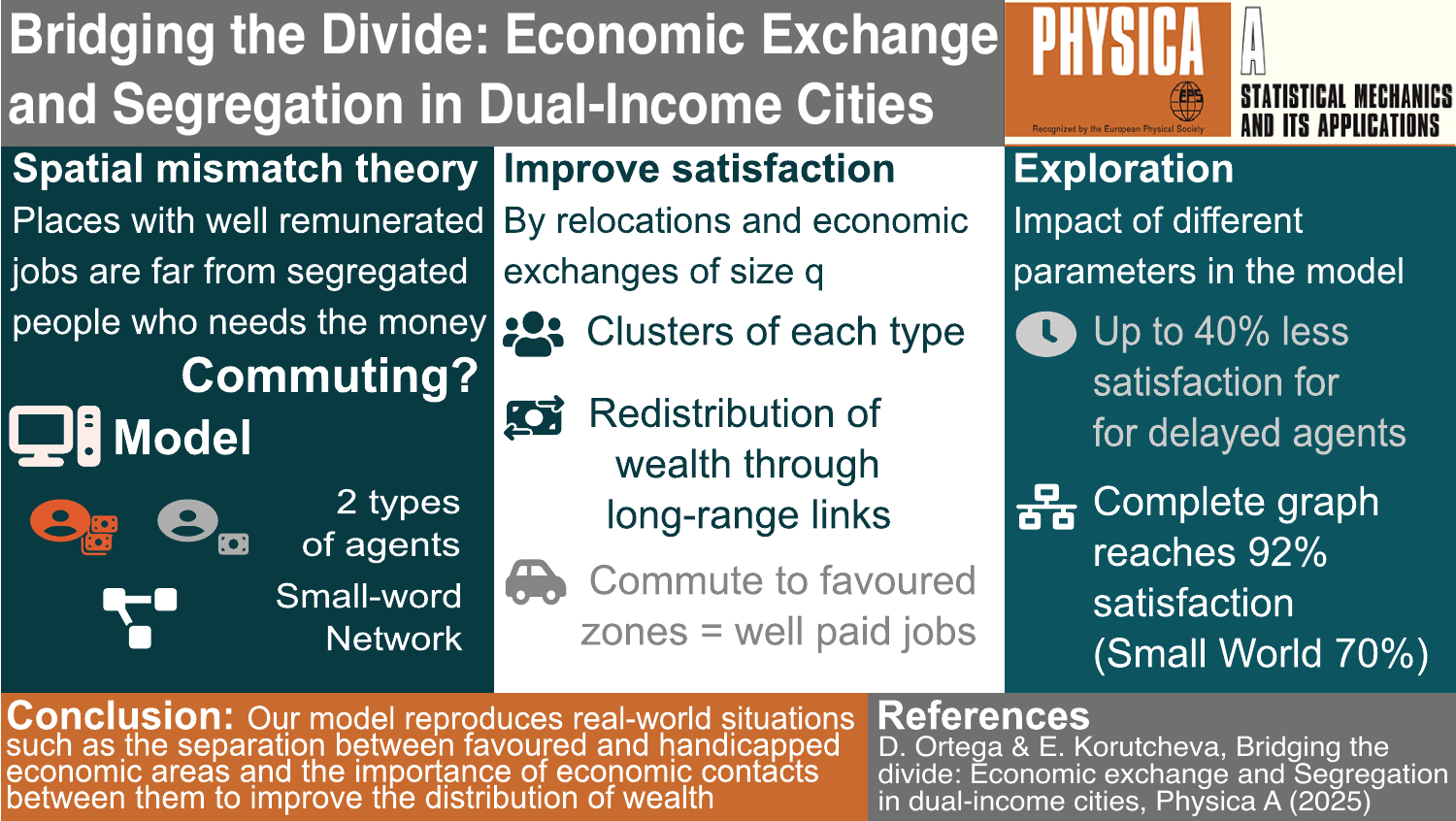}
\end{graphicalabstract}

\begin{highlights}
\item Extension of the Schelling model with economic terms over a small-world network
\item Financial exchanges occur exponentially, but relocations are modeled with power laws
\item The size of the economic exchange plays a crucial role
\item Delayed agents are also considered
\item Highlights how network links influence wealth redistribution
\end{highlights}


\begin{keyword}
Sociophysics \sep Segregation \sep Schelling's model \sep Commuting  \sep SW Networks



\end{keyword}

\end{frontmatter}



\section{Introduction}
\label{sec:intro}

Segregation, the separation of groups based on differing characteristics, frequently implies unequal treatment \cite{Britannica(2025)}.  This division can occur along various lines, including religion, race, culture, and socioeconomic status \cite{Yao(2018)}. The agglomeration of these discriminated groups in specific urban locales has resulted in the genesis of ghettos. These zones are characterized by a concentration of poverty, limited educational opportunities and high exposure to crime and violence \cite{Massey(1990)}. These conditions may engender a \textit{poverty trap}, which is defined by a self-reinforcing mechanism that leads to increased poverty in the absence of external intervention \cite{Azariadis(2005)}. 

Spatial segregation is characterized by five dimensions: evenness, exposure-isolation, concentration, centralization and clustering \cite{Massey(1988)}. Basically,  results support the \textit{spatial mismatch theory}: high-income areas which can offer well-remunerated jobs are distant from the most deprived zones characterized by high unemployment rates \cite{Massey(1988), Duncan(1955), Simkus(1978), Gobillon(2007)}. Consequently, the poverty traps previously referenced have a tendency to persist over an extended period of time. The most efficient solution to this predicament is to seek employment in distant locations, wich gives rise to long daily commutes.

Segregation in mobility patterns have been recently studied, finding that residents of White neighborhoods are far less likely to visit non-white neighborhoods, while urban dwellers of black and Hispanic neighborhoods visit White areas slightly less than they visit other similar ones \cite{Vachuska(2023)}. As we can infer, Black–White residential segregation is a primary predictor of segregated
urban mobility patterns in the largest 50 cities in the USA \cite{Candipan(2021)}. Likewise, characterizing the relation between socioeconomic status and mobility for US cities, results shown that low-income populations tend to make shorter trips, but more frequents \cite{Barbosa(2021)}. Another study highlights that although commuter mean time for black people is usually longer than that of white people, it is converging in medium and small-sized cities during the last years, due to the use of the car as a means of transportation \cite{Bunten(2024)}. It has been observed that the mobility networks within these cities tend to exacerbate social segregation, owing to the presence of a wide variety of spaces that are designated for different socio-economic strata in large cities\cite{Nilforoshan(2023)}. The dynamic relation between mobility patterns and economic inequalities which gives rise to segregation is analyzed via mobile phone data on an individual level in a recent contribution\cite{Moro(2021)}. 

Following a thorough analysis of relevant literature from a socio-economic standpoint, this study will commence its theoretical investigation into models of segregation. The primary study to be analysed is the Schelling model, wherein two groups of agents are situated on a grid. These agents have the capacity to relocate to an unoccupied position, provided that doing so would render them \textit{happy} or \textit{satisfied}; in other words, the vicinity composition of their new location must overcome a treshold value linked to the number of different and equal neighbors \cite{Schelling(1971)}. A housing market and different tolerance values for black and white people were taken into account in \cite{Zhang(2004)}, finding that real estate plays an important role in the dynamics of segregation, given that people in the economic favoured group bid up housing prices, hence keeping away people from the other group. Another interesting contribution was able to predict the relocation of higher-status households in suburban zones by using a cellular-automata \cite{Fossett(2006)}. The real estate market of Paris was modelled considering the propensity of each prospective purchaser to procure a given property is contingent, in part, on the inherent appeal of the property itself, and, in part, on the social characteristics of the neighbourhood \cite{Gauvin(2012)}.

Recent research has seen an increased focus on complex agent-based models incorporating ethical considerations. Studies have explored the significant impact of altruistic agents on system outcomes and overall well-being \cite{Jensen(2018)}. The influence of both altruistic and fair agents has been evaluated in \cite{Flaig(2019)}, while the behaviour of agents with adaptive tolerance to their surroundings has been examined in \cite{Urselmans(2018)}. Open city simulations, allowing agents to enter and exit the city, have been employed to model gentrification processes \cite{Ortega(2021)}. Nonetheless, it should be noted that these models predominantly utilise the classical rectangular lattice, thereby limiting their applicability to real-world urban environments.

In the context of our research, several studies employing the Schelling model on networks are of particular pertinence. The influence of moderate tolerance preferences on segregation in both lattice and network structures is compared in \cite{Fagiolo(2007)}. This study also demonstrated that polarization mechanisms occur not only in regular spatial networks but also in more general social networks. The influence on the values of various segregation indices has been characterized across different network topologies, finding that the system evolves towards steady states exhibiting maximum segregation \cite{Cortez(2015)}. In contrast to previous studies, which found similar results despite inherent network differences, the importance of \textit{cliques}, clusters that reinforce segregation, is highlighted in \cite{Banos(2010)}. Consequently, this work offers suggestions for addressing real-world urban situations that can generate clique-like structures. The economic segregation experienced by individuals was accurately predicted by adding an Schelling extension to a classic model of individual mobility \cite{Moro(2021)}. At the scale of census tracts, an extended Schelling model including a real state market over networks was able to classify areas as ghettos and non-ghettos with $80\%$ accuracy for the city of Washington D.C. \cite{Ortega(2022)}

While the Schelling model and its variants offer insights into segregation, it does not fully capture the complexities of its link to commuting. In order to bridge this gap, we have developed an extended Schelling model with a simple economic exchange framework and a real estate market over small-world networks. We aim to understand how commuting segregation arises and how its effects might be minimized. Our model, though simple, effectively replicates real-world phenomena.  These include the division between prosperous and economically disadvantaged areas, the crucial role played by economic interaction between these areas in distributing wealth through commuting, and the importance of affordable, efficient transportation to disrupt poverty cycles in struggling communities.

The paper is organized as follows. In Section \ref{sec:model} we define the agent \textit{happiness} or \textit{satisfaction} level, defined as $I_{dis}$, the system dynamics and the considered network. In Section \ref{sec:results} we characterize the system evolution and its final state. Furthermore, our study examined the variability in agent satisfaction when the parameters of the system were modified and across various networks, analysing the relationship with commuting behaviours. Finally, our main conclusions and proposals for further work can be found in Section \ref{sec:conclusion}.

\section{Model}
\label{sec:model}
\subsection{Network}

In selecting a network to illustrate urban transportation systems, two pivotal factors must be considered. Firstly, segregation has significantly influenced the urban layout, with the presence of barriers that impede the integration of marginal neighbourhoods being especially prominent \cite{Toth(2021)}. This suggests a propensity to congregate areas that are analogous in economic status. Conversely, the city must be interconnected, requiring communication pathways between disparate regions, thus ensuring the absence of isolated areas excluded from the city \cite{Bettencourt(2013)}.

\begin{figure}[H]
\center
\includegraphics[width=12 cm]{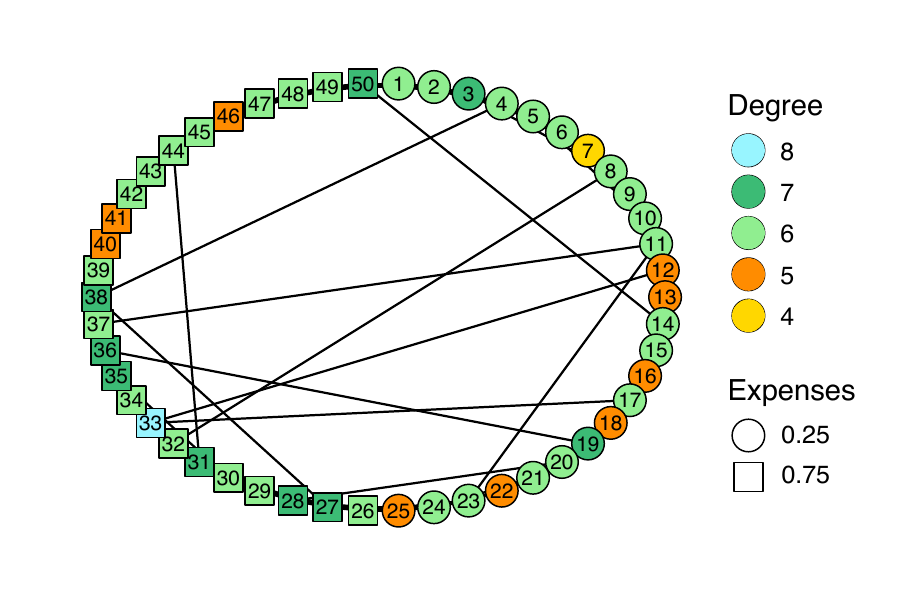}

\caption{Small-world network of $50$ nodes with $6$ nearest neighbors and a rewiring prpbability of $0.1$. Two possible values of housing expenses are illustrated via node shapes. Node degrees are associated to the color scheme. The represented network will be considered fixed unless otherwise stated.}
\label{Fig:sw}
\end{figure}

The desired balance between the two factors can be achieved by implementing a small-world type network on which a real estate market with two disparate values is defined, as it is illustrated in Fig.\ref{Fig:sw}. Our framework consists of $50$ nodes, with links generated considering a low rewiring probability of $p=0.1$ \cite{Watts(1998)}. The number of nearest neighbors considered is $6$. When calculating the census tracts bordering a given one for Washington D.C., an archetypal case of a segregated city that we adopted as a reference, an average value of $6.01 \pm 1.95$ is obtained \cite{Ortega(2022)}.

\unskip

As can be seen in Fig. \ref{Fig:sw}, positions $1$ to $25$ were assigned a low housing expense, $E=0.25$ arbitrary economic units. Conversely, the remaining nodes are considered  expensive, with a value of $E=0.75$. Consequently, within this network, we observe that zones with similar socio-economic levels are strongly connected, while only a limited number of long-distance connections link deprived and economically favored areas

\subsection{Dissatisfaction Index}

Once our network is established, we focus our attention on the agents. 
We use a model inspired by the Schelling's work\cite{Schelling(1971)}, featuring \textit{red} and \textit{blue} agents distributed across the previously defined network of nodes, some of which may be vacant. An agent's tolerance for diversity is quantified by the fraction of dissimilar neighbors, $f_d$, calculated as $N_d/(N_s+N_d)$, where $N_s$ and $N_d$ represent the number of similar and dissimilar agents in adjacent nodes, respectively. To calculate $f_d$, we restrict our neighborhood consideration to the six nearest nodes, omitting far-off connections, and assuming that the concept of neighborhood, related to tolerance, refers to those with physical proximity. Unlike traditional lattice models with fixed neighborhood sizes, our network allows for variable neighbor counts per node. 

Furthermore, the model considers housing with varying economic levels based on network location. To finance this housing, the model assigns varying wages to agents, and economic exchanges related to commuting occur between them. Although a wide range of financial values have been studied, we assume that red agents possess sufficient economic resources to cover their housing expenses having a surplus, $W_r \backsim \mathcal{U}(0.75,1.0)$, whereas blue ones require economic support to cover their living costs, $W_b \backsim \mathcal{U}(0,0.25)$. Therefore, they need to commute to the financial favoured zones to earn a salary.  It must be noted that economic exchanges are measured in arbitrary units with a fixed value $q$, and can be performed between all adjacent nodes. In other words, distant nodes are also included.

	An agent's state is assessed using the dissatisfaction index, $I_{dis}$ defined as:

\begin{equation}
	I_{dis} = f_d - T + [W_i - E(n) - T_e],
	\label{eq:Idis}
\end{equation}
 
where $T$ is the tolerance treshold, the maximum value of $f_d$ that an agent can withstand while remaining \textit{happy} discarding other factors. Terms in the bracket are an extension of the segregation model which includes the wages of the agent $i$ depending on her/his type,  $W_i$, and the housing expenses associated to the $n$ position of the network $E(n)$. Finally, $T_e$ is the economical tolerance to the difference between the income of the agent and the housing costs of the occupied node. 

It must be noted that lower (negative) values of $I_{dis}$ in Eq.\ref{eq:Idis},  imply greater \textit{happiness} and vice versa. Regarding segregation, an agent attains maximum happiness, $-T$, when all of its immediate neighbours are of a similar type, $f_d=0$. From an economic perspective, the optimal value of $-T_e$ is reached when income and expenditure are precisely equal, $W_i=E(n)$. This reflects the fact that people adapt their expenses to their salary level. However, if an agent's expenses exceed their income, $W_i - E(n) < 0$, they experience a decline in happiness irrespective of the characteristics of their immediate neighborhood, which is to say $I_{dis} > 0$. Consequently, red agents can offer monetary compensation to blue agents for their work, thereby increasing satisfaction up to the point where further economic exchange becomes detrimental. As a consequence, the economic unit $q$ plays a major role in system dynamics. From red agents' perspective, low values of $q$ foster wealth distribution by narrowing the wage-expense gap, whereas high values of $q$ restrict exchanges to red agents with substantial monetary surpluses.

\subsection{System dynamics}
We start from a random initial configuration with equal proportions of red and blue
agents and a small percentage of vacancies, $8\%$. This low percentage correlates with the high population density within cities. As we consider a closed city model the same number of vacancies remain during the simulation.

At each iteration, we choose an unhappy agent and consider a relocation or economic exchange with equal probabilities. For relocations, a vacany is randomly selected. If the exchange verifies that the agent is happier in the vacancy, i.e. lower dissatisfaction index $I_{dis}$, relocation takes place leaving empty the previous occupied position. Otherwise, the move is rejected. For economic exchanges, an agent who needs money must find a neighbor, randomly selected from the adjacent nodes, who has a surplus and is willing to pay them. Now, two conditions must be met: first, people usually strive to balance their income and expenditures, and consequently, their happiness tends to increase as the gap between these two narrows (see Eq. \ref{eq:Idis}), assuming they avoid accumulating debt $W_i - E(n) \geq 0$. Second, the agent who works must need the money $W_i - E(n) < 0$. In this way, both agents improve their happiness due to the transaction. This iteration is repeated until the system reaches an equilibrium state where all changes are discarded. We define a \textit{Monte-Carlo} step as an iteration that runs over all the unhappy agents.

\section{Results}
\label{sec:results}

This section is organized into two distinct parts, each comprising three subsections. The first part provides a foundational analysis of the system, exploring its social (\ref{subsec:social}), economic (\ref{subsec:economy}), and evolutionary dimensions (\ref{subsec:evol}). Building upon this foundation, the second part investigates the correlative relationships between agents' final state and three key variables: the parameter $q$ (\ref{subsec:qs}), action delays (\ref{subsec:delays}), and network topology (\ref{subsec:networks}).

For the socioeconomic and evolutionary part, results are obtained from a number of runs $N_R=10,000$ over the network previously depicted in Fig. \ref{Fig:sw}. The economic exchange size is  $q=0.005$, assuming that small economic exchanges are possible. While the model incorporates a moderate tolerance level for neighborhood diversity of $T=0.5$ (agents are \textit{happy} with at least $50\%$ similar neighbors), economic factors play a more substantial role in determining overall satisfaction. We posit that individuals possess greater agency in managing their finances, and therefore, the inability to meet a balance between wages and expenses results in a significantly higher degree of unhappiness compared to living alongside dissimilar agents, thus $T_e=0.1$. A stark contrast exists between the effect of $T_e$ over the agents: for the blue ones, this value is irrelevant as a positive salary-expense balance is what they seek. For affluent agents, red ones, economic terms contribute to satisfaction if $W - E \leq 0.1$, thus acting as a threshold of the monetary balance. It may also be understood that wealthy agents whose spending is restricted beyond a limit in their salary-expense balance experience a source of unhappiness. 

\subsection{Population distribution and happiness}
\label{subsec:social}

The main question about the Schelling model relates to the position and type of agent. To address this, a spin has been linked to the node in $i-th$ position, $S_i$\cite{Blume(1971), Gauvin(2010),Ortega(2021a)}. With a red agent occupying the spot, $S_i=+1$, whereas a blue agent implies $S_i=-1$ . If node is empty, $S_i=0$. Thus, position i's mean spin, is a sum over all the spin values of the place \textit{i} normalized by the total number of runs, i.e. $\bar{S_i}=1/N_R\sum_{k=1}^{N_R}{S_{i,k}}$.  Values $\bar{S_i} \approx  \pm {1}$ accurately reflect high probabilities of red and blue agent occupancy, respectively. However, states with $\bar{S_i} \approx 0$ are susceptible to misinterpretation. These values may arise from either equal attraction to both agent types or a high probability of vacancy. To provide a more robust measure, we define the average occupation of the node \textit{i}, $\bar{O_i}$, as the number of times that node \textit{i} is filled by an agent divided by $N_R$. The definition of average happiness, $\bar{H_i}$ is analogous to that of occupation, but considering happy agents in each node \textit{i}.

\begin{figure}[H]
\center
\includegraphics[width=12 cm]{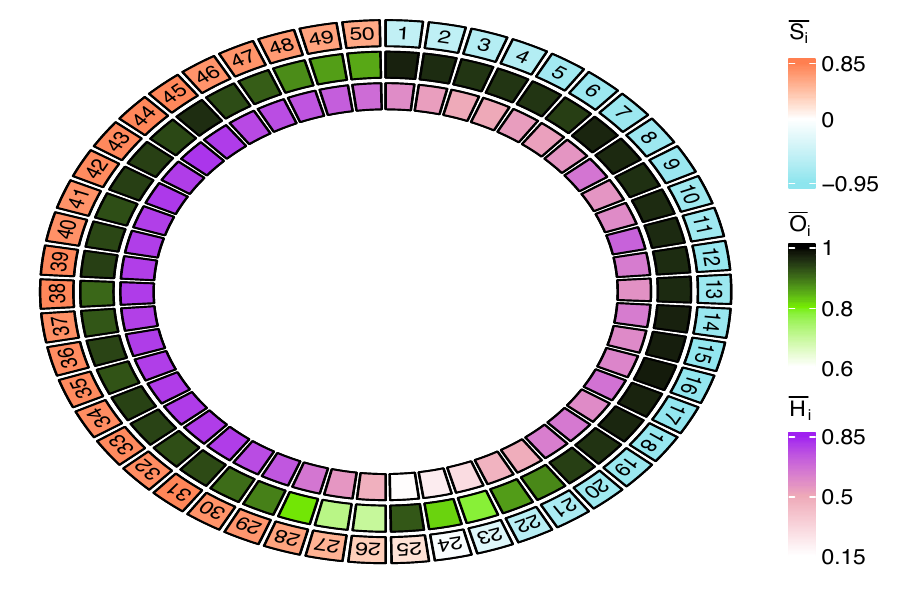}
\caption{Circular graph where each sector corresponds to the node in the system from Fig. \ref{Fig:sw}. $\bar{S_i}$, $\bar{O_i}$ and $\bar{H_i}$ denote the the mean spin, occupation and happiness in the $i$ node.}
\label{Fig:rueda1}
\end{figure}

These coefficients, $\bar{S_i}$, $\bar{O_i}$ and $\bar{H_i}$, have been simultaneously depicted and associated to a color scheme in Fig. \ref{Fig:rueda1}. A significant spatial segregation was observed in the outer circle, with blue agent clusters predominating in nodes $2$ to $23$, whereas red agents cluster on the opposing side, from $27$ to $47$. Segregation is likely driven by homophily and modulated by economic constraints. It must be noted that the intermediate circle demonstrated a marked decrease in agent occupancy at the cluster interface, brighter tones from $26$ to $28$ and $48$ to $50$, a phenomenon consistent with prior research on vacancy formation \cite{Gauvin(2010), Ortega(2021a), Ortega(2021b)}. Despite moderate tolerance parameters, agents exhibited a preference for intra-cluster positioning, resulting in reduced occupancy at the border. 

\unskip

\vspace{6pt}
Finally, if we look at the innermost circle of Fig. \ref{Fig:rueda1}, we observe that red agents displayed enhanced happiness in regions with higher property values. Conversely, blue agents exhibited lower satisfaction values and a preference for cheaper market regions. This pattern bears resemblance to historical \textit{redlining} practices, where access to affordable housing was systematically restricted for minority populations, often relegating them to peripheral or economically disadvantaged areas\cite{Rothstein(2018)}. The probability of being happy for each type of agent is calculated as the ratio of satisified agents to their total number ($23$) averaged over $N_R$ iterations. For a red agent is $0.82\pm0.15$, a high value attributed to their strong financial standing and clusterization. The measure decreases to $0.58\pm0.11$ for blue agents, which have a monetary handicap. However, it is noteworthy that certain locations ($8,11,12,17,19$ and $20$) exhibit a higher level of satisfaction than their surrounding nodes. This may be attributed to economic exchanges, which we explore in further detail below.

\subsection{Economic exchanges and commuting}
\label{subsec:economy}
As established, red agents possess financial capital which they can allocate to blue agents, who demonstrate a deficit. The mechanism facilitates the attainment of a satisfaction state for both agent types, as formalized by the Eq. \ref{eq:Idis}. Therefore, there exists a transfer of money between nodes in the form of monetary units $q$. To study the net economic transfers, we define a matrix $\textbf{T}$. In this matrix, when node \textit{i} pays one \textit{q} to node \textit{j}, we subtract one unit from the element $T_{ij}$. Conversely, when node \textit{i} receives \textit{q} from node \textit{j}, we increase the matrix element $T_{ij}$ by one unit. This ensures the conservation of the total economic resources within the system being $\textbf{T}$ antisymmetric. 

\begin{figure}[H]
\center
\begin{tabular}{cc}
\includegraphics[width=7cm]{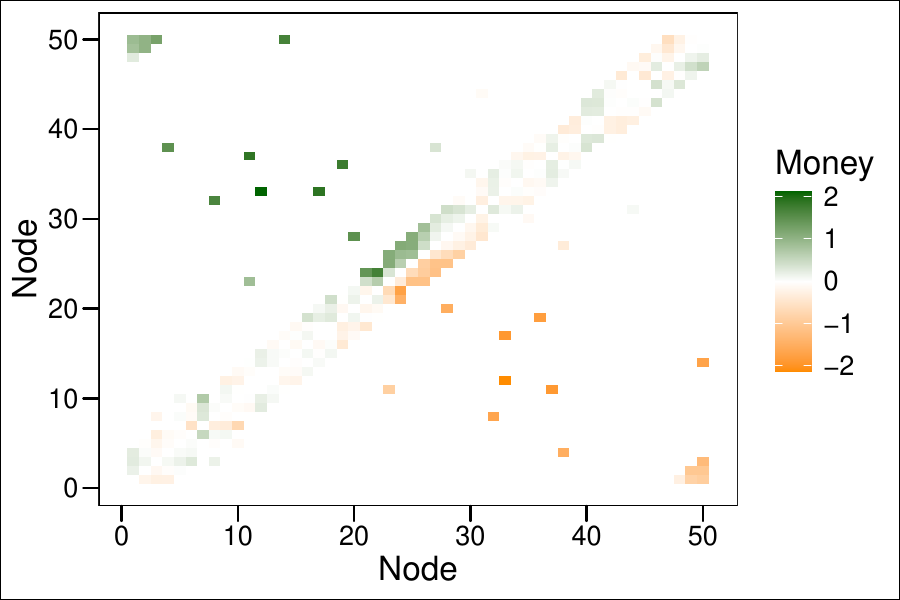} &
\includegraphics[width=7cm]{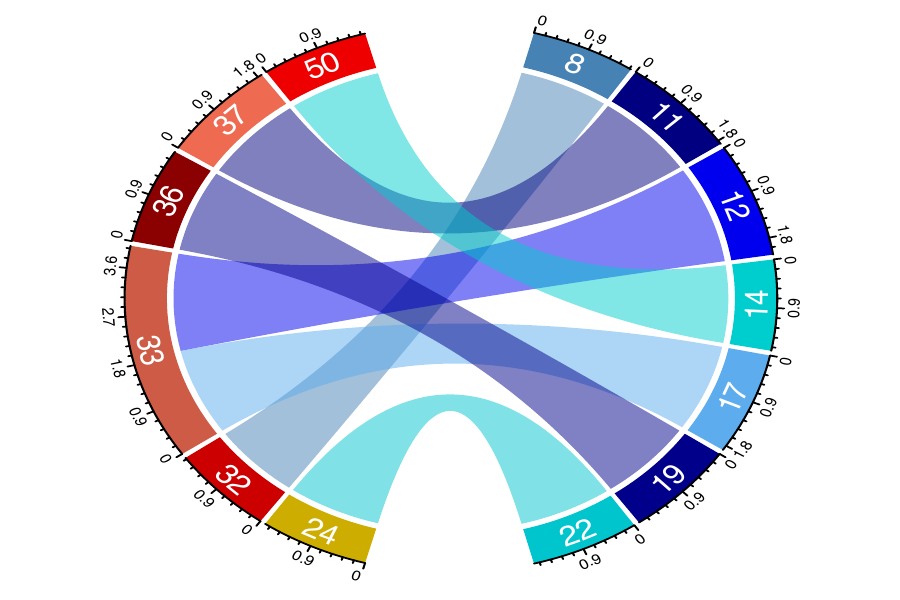}\\
\textbf{(a)} & \textbf{(b)} \\
\end{tabular}
\caption{(\textbf{a}) Economic exchange matrix $\textbf{T}$ in $q$ monetary units. Positive values indicate that the node acts as a monetary sink, while negative values indicate sources. The matrix is antisymmetric, given that when one agent gives a quantity $q$, $-1$ for the agent, the other receives it, $+1$. (\textbf{b}) Chord diagram of economic exchanges exceeding $|T_{ij}| \geq 1.5$, where $i$ and $j$ are node numbers. The value was chosen to preserve the principal net economic transfers between regions, see (\textbf{a}) scale. The external number indicates the mean number of exchanges.}
\label{Fig:money}
\end{figure}

The average value of this matrix over $N_R$ runs is represented in Fig. \ref{Fig:money} \textbf{a)}. As it can be seen, dark yellow squares, indicative of primary financial resource donors, exhibits a positive correlation with higher node numbers (x-axis values), corresponding to the places occupied preferently by red agents. Conversely, the primary recipients of these resources are concentrated at lower node numbers, aligning with the right segment occupied by financially disadvantaged blue agents, as it was depicted in Fig. \ref{Fig:rueda1}. These exchanges occurs between adjacent nodes linked by long-range connections between economically advantaged and handicapped regions. Notably, intra-agent type financial transactions are absent in net value. This may be due to the economic distributions: red agents, while possessing financial resources, do not require them, and blue agents lack the capital to compensate neighboring agents. Consequently, mobility deprived zones to favoured regions is needed, giving rise to a commuting framework. Nevertheless, there are two zones which coincide with the border between clusters, where these transactions can also occur via short-range links.

Retaining the elements of Fig. \ref{Fig:money}\textbf{a)} whose absolute value exceed the threshold of $1.5$, we obtain Fig. \ref{Fig:money}\textbf{b)}. In this figure, the colour of the outer area indicates whether this area is frequently assigned to blue agents (on the right) or red agents (on the left). Furthermore, it allows us to establish an association to the previous section \ref{subsec:social}, as we observe that the most satisfied blue agents from Fig. \ref{Fig:rueda1} are those with links to economically favoured areas. Marginal neighbourhoods with good connections that facilitate commuting, have access to greater economic resources increasing their satisfaction level. On the other hand, economically deprived zones which are isolated from richer regions exhibit lower levels of happiness.

The role of node $24$, depicted in yellow in Fig. \ref{Fig:money}\textbf{b)}, is noteworthy. This area exhibits an approximately null spin value yet is connected to node $22$, which is predominantly blue. Consequently, it serves as an economic source for this region. Additionally, it is important to consider that the degree of the nodes influences the economic balance established between them. Given the system's dynamics, a node in need of economic resources will search a monetary improvement between neighbors with equal probability. As a result, nodes with a lower degree, such as the node $22$ with $5$ neighbors, tend to acquire more average resources compared to those with a higher degree ($24$ has $6$), increasing the net monetary flux towards them.

\subsection{System evolution}
\label{subsec:evol}
There are two types of movements that agents can perform to improve their satisfaction: economic exchanges and relocation to another node. We now examine how these changes evolve considering the entire system. 

\begin{figure}[H]
\center
\begin{tabular}{cc}
\includegraphics[width=7cm]{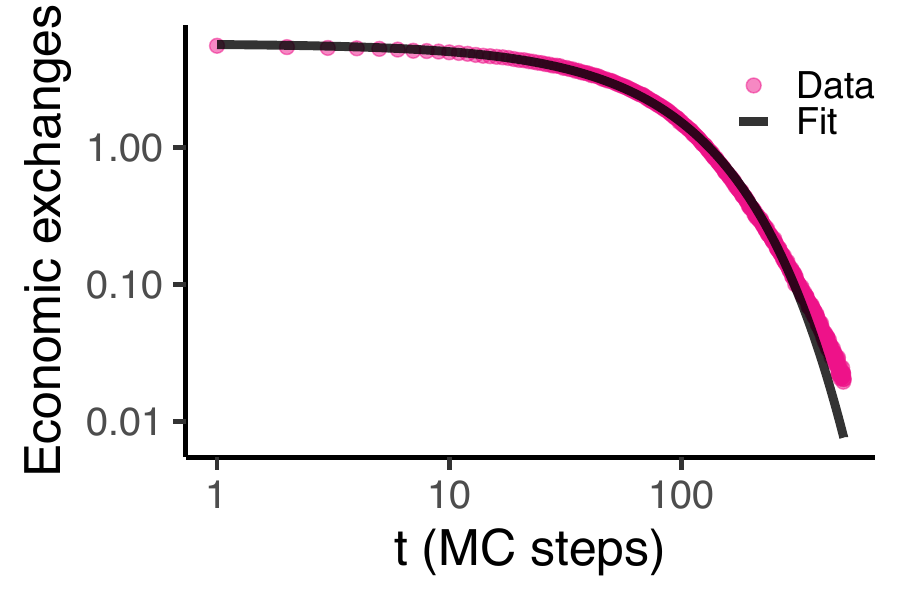} &
\includegraphics[width=7cm]{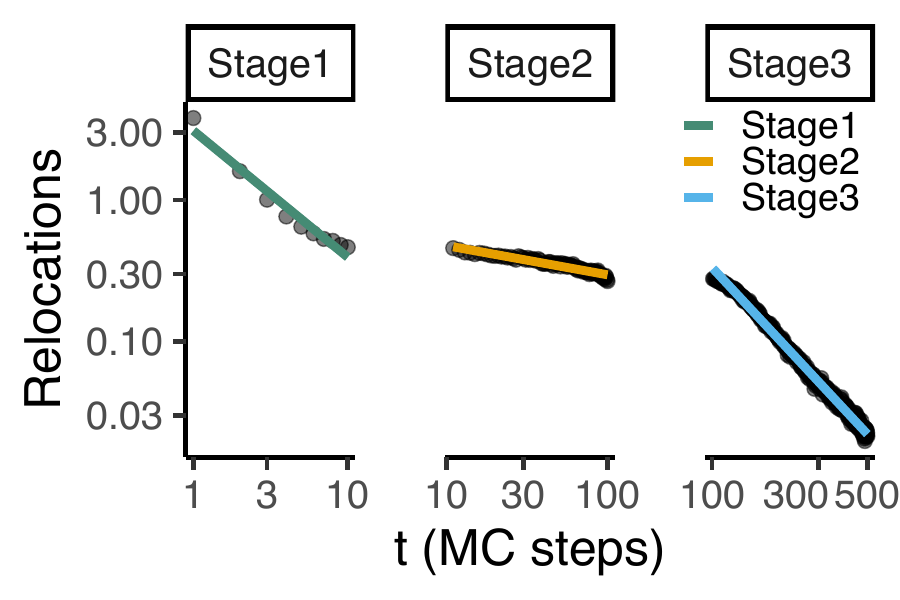}\\
\textbf{(a)} & \textbf{(b)} \\
\end{tabular}
\caption{Evolution of the number of economic exchanges (\textbf{a}) and relocations (\textbf{b}). Time is measured in Monte-Carlo (MC) steps. Each step involves a relocation or economic exchange attempt of all the unsatisfied agents in the system. Measures are averaged over 10,000 system runs.}
\label{Fig:evol}
\end{figure}

Economic exchanges follow a decreasing exponential law, as depicted in Fig. \ref{Fig:evol}\textbf{a)}. The reason for this is the redistribution of wealth as a diffusion process. Red agents in expensive housing areas have an excess of money given by $\mathcal{U}(0,0.25)$; blue agents, even when located in cheap regions, have an economic need described by $\mathcal{U}(-0.25,0)$. Since we fix exchanges of size $q$, we can understand that red agents act as sources and blue agents as sinks of money. That is, there is a concentration of units $q$ in wealthy agents that tends towards regions that absorb this excess. These exchanges of $q$ continue as long as both benefit: red agents do not go into debt by donating money, and blue agents stop earning it once they have reached happiness, being $W_i - E(n) \geq 0$.

It should be emphasised that, unlike economic exchange, relocation takes place in three distinct stages, described by power laws as it is depicted in \ref{Fig:evol}\textbf{b)}. Initially, in the stage 1, there are vacancies in locations of interest to the agents, leading to numerous changes until these vacancies become unattractive. During the stage 2, the probability of any relocation within the system is low, but not zero. This is due to occasional changes within the system as a result of economic exchanges, which continues at an almost constant rate during this period. Finally, in stage 3, the probability of relocation drops sharply, suggesting that the system has reached a state of equilibrium in the previous period.

The interplay between these processes resembles what happens socially: Once individuals have chosen their preferred residential area, taking into account their socio-economic level, they try to adjust the balance between income and expenditure in order to increase their satisfaction.

\subsection{Economic exchange sizes}
\label{subsec:qs}
\begin{figure}[H]
\center
\begin{tabular}{cc}
\includegraphics[width=7cm]{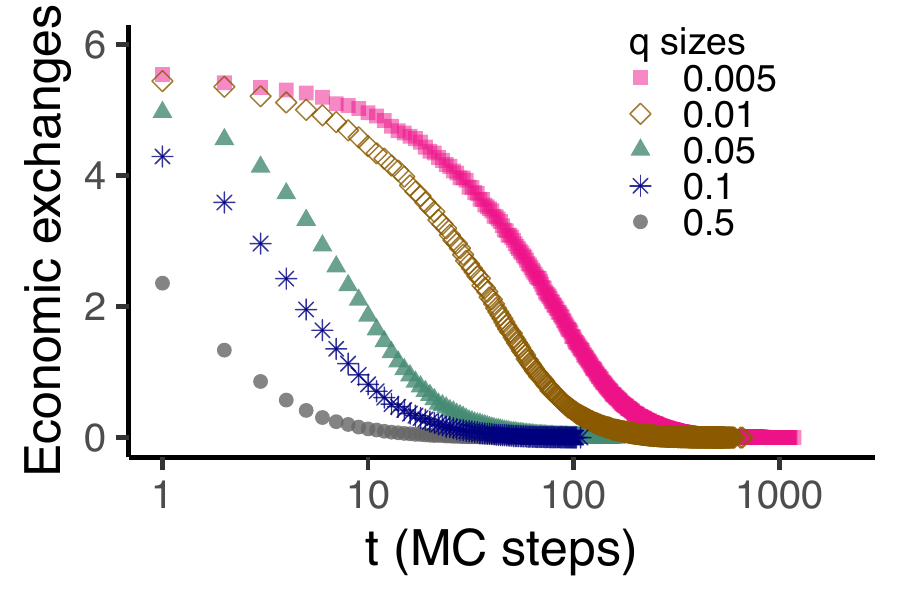} &
\includegraphics[width=7cm]{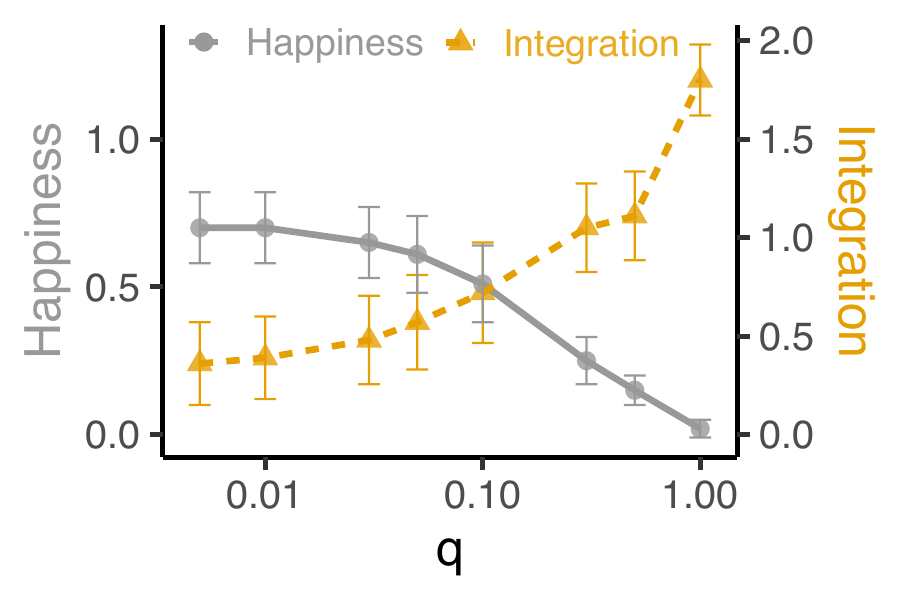}\\
\textbf{(a)} & \textbf{(b)} \\
\end{tabular}
\caption{ (\textbf{a}) Economic exchanges evolution in Monte-Carlo steps for different economic exchange sizes $q$. (\textbf{b}) Mean agent happiness (grey) and Integration (gold) versus $q$.}
\label{Fig:qstate}
\end{figure}

Now, we focus our attention to the influence of the discrete economic unit $q$ on the dynamics of economic exchanges. The significance of $q$ lies in its dual role: an elevated $q$ accelerates wealth redistribution by increasing the magnitude of individual transactions, facilitating rapid transfers of monetary value, as it can be observed in Fig. \ref{Fig:qstate} \textbf{a)}. Conversely, an increase in $q$ diminishes the population of agents possessing sufficient liquidity to efficiently adapt their income-expenditure balance. This reduction in adaptable agents impedes the diffusion of resources between source and sink nodes within the economic network, potentially leading to inefficiencies in resource allocation.

Nonetheless, $q$ also plays a pivotal role in the integration and satisfaction of the entire system, as depicted in Fig. \ref{Fig:qstate}\textbf{b)}. Average agent happiness $\bar{H}$ is defined as the fraction of happy agents normalized by their total number, $N_A$. For integration, we use the measure $C_I=L_d/N_A$ \cite{Banos(2010)}, where $L_d$ is given by the number of edges which link different type of agents. 
When $q$ is small, economic exchanges are efficient, and $C_I$ is low, suggesting the formation of two distinct clusters of opposing agents. This segregation, paradoxically, facilitates efficient long-range commutes, enabling wealth redistribution and a high happiness value $\bar{S}$. When $q$ increases, the number of economic exchanges decreases, as wealthy areas are unwilling to cede monetary amounts that would lead them into debt. Since the system begins with a random distribution and few vacancies are available, as occurs in densely populated cities, few agents can achieve happiness simply by switching locations. Thus, for large $q$, both integration and dissatisfaction increase, as wealth redistribution fails.

Consider a scenario in which an economic agent incurs significant temporal or financial expenditure during transit to secure an economic benefit of magnitude $q$. If the value of $q$ is insufficient, the existing inefficiencies in commuting are likely to preclude the occurrence of this transaction. Consequently, the viability of such an exchange is contingent upon either an increase in the economic gain $q$, representing a higher remuneration, or a substantive improvement in the efficiency of the transportation infrastructure.

This phenomenon aligns with the \textit{spatial mismatch hypothesis}, which postulates that financial limitations necessitate extended commutes for individuals seeking employment opportunities in affluent locations. This underscores the policy relevance of investing in accessible and cost-effective transportation systems to mitigate spatial disparities and foster development in disadvantaged communities.

\subsection{Delays}
\label{subsec:delays}

\begin{figure}[H]
\center
\begin{tabular}{cc}
\includegraphics[width=7cm]{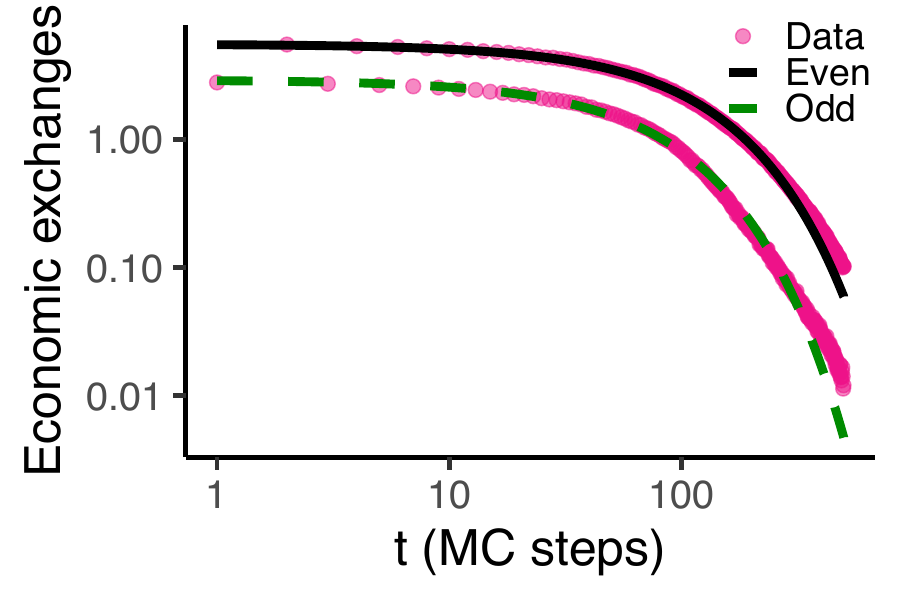} &
\includegraphics[width=7cm]{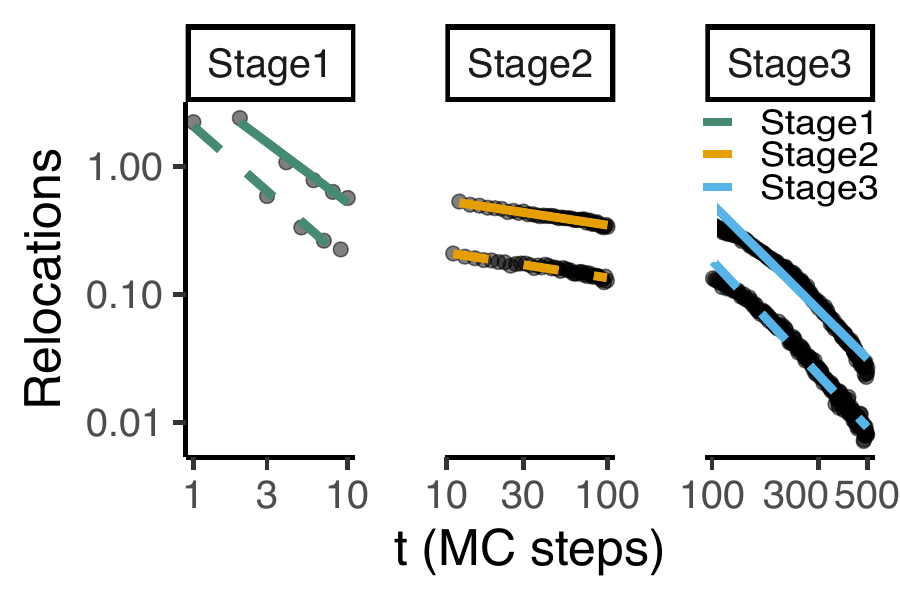}\\
\textbf{(a)} & \textbf{(b)} \\
\end{tabular}
\caption{Evolution of \textbf{the number of} economic exchanges (\textbf{a}) and relocations (\textbf{b}) considering half the agents of both types take two steps to finish an action. The terms 'odd' and 'even' of (\textbf{a}) points out the temporal parity of the step value. Hence, continuous lines depict the movement of all agents, while dashed lines are associated with non-delayed agents. Whereas economic exchanges appear to follow a consistent trend throughout their evolution, relocations (\textbf{b}) exhibit three highly distinct stages, all of which are amenable to power law fitting. Measures are averaged over 10,000 systems runs.}
\label{Fig:delay2}
\end{figure}

Segregation models typically assume an immediate response: agents decide to relocate to a new position within the same time step. However, in reality, household relocation is a time-consuming process due to the multitude of factors involved, including real estate market prices, salaries, and routine household expenses. To evaluate the system's response to this delay, we assume that, statistically, both half of the red agents and half of the blue agents require two time steps to either change position or complete an economic exchange. In other words, each agent has a $50\%$ of being delayed.System evolution is shown in Fig. \ref{Fig:delay2}.

To understand the depicted dynamics in Fig. \ref{Fig:delay2}, it is crucial to note the distinct operational frequencies of the agents within the model. Specifically, half of the agents execute their actions with every discrete time step, while the remaining half operate at a one-step delay, updating their states only every two time steps. Consequently, during even-numbered time steps, all agents in the system are active, contributing to a higher observed rate of changes. During odd-numbered time steps, only half the agents update their states. Fig. \ref{Fig:delay2} \textbf{a)} illustrates the impact of temporal delay on economic exchange dynamics. Specifically, the data reveal two exponential decay curves exhibiting a consistent downward trend. This phenomenon is also observed in agent relocation processes, as depicted in Fig. \ref{Fig:delay2} \textbf{b)}, pointing out a consistent effect of temporal delay across different system interactions. The delay's influence on relocations and economic exchanges manifests distinctly. While the system's vacancy count remains constant, relocations are only accessible to non-delayed agents in the subsequent odd time step. Conversely, for economic exchanges, which occur more slowly and randomly among all adjacent nodes, the impact is less severe because neighboring agents typically preserve their cluster configurations.

Although the system evolution provides insight into the relationship between agents with and without delays, the primary focus of this analysis is how the total time of delay affects their satisfaction. Time delays measured in MC steps of $\delta=2,5$ and $10$ were considered. These data have been compared to those of a system without delayed agents, being the results in Table \ref{Tab:delays}. For example, in the first column, with $\delta=2$, the value of $0.33\pm 0.10$ indicates the probability of happiness for blue agents without delay, while the value of $0.23\pm 0.10$ is for blue agents who take $2$ steps to finish their actions (D). The last column of the table, Non-delayed, refers to the system where no agent has a delay, i.e. the usual system. Values demonstrate that agents with higher financial resources are more resilient to satisfaction decline than those with limited funds. This is due to the difference in satisfaction drivers: red agents rely on neighborhood composition (Equation \ref{eq:Idis}), while blue agents need to manage finances. Access to well-connected networks is pivotal for blue agent. However, because relocations occur quickly (Subsection \ref{Fig:evol}), even a momentary delay can prevent them from reaching these key nodes and negatively affect their happpiness. In fact, as the delay increases, the blue delayed agents' satisfaction decreases.

\begin{table}[H]
\begin{center}
\begin{tabular}{|c|c|c|c|c|}
\hline
\textbf{Agent}	& \textbf{$\delta=2$} & \textbf{$\delta=5$} & \textbf{$\delta=10$}	& \textbf{Non-delayed}\\
\hline 
\hline 
Blue 		& $0.33\pm 0.10$	&	$0.36\pm 0.10$ &$0.38\pm 0.10$	& $0.57\pm 0.11$\\
\hline 
Blue (D)		& $0.23\pm 0.10$	&$0.20\pm 0.10$	&$0.16\pm 0.10$& 	 \\
\hline 
Red 		& $0.42\pm 0.11$	&$0.43\pm 0.11$	&$0.41\pm 0.12$	& $0.81\pm 0.15$\\
\hline 
Red (D)		& $0.39\pm 0.13$	&$0.36\pm 0.13$	&$0.31\pm 0.13$	 & \\
\hline 
\end{tabular}

\par\end{center}
\caption{Fraction of satisfied agents of each color, assuming a $50\%$ likelihood of being delayed. This involve that, statistically, half of them require $\delta$ Monte Carlo steps to complete the action. Delayed agents are identified by the letter (D). The reference system, in which no agents experience a delay ($\delta=1$), corresponds to the 'Non-delayed' column.}\label{Tab:delays}
\end{table}

Table \ref{Tab:delays} shows that overall satisfaction levels were similar between the systems, even when delays were introduced. This observation demonstrates the influence of the connectivity of the blue agents at a limited number of nodes over the overall satisfaction, thus underscoring the main role played by the network structure.

\subsection{Different Networks}
\label{subsec:networks}

We now proceed to analyze the remaining fixed parameters to ascertain the network configurations that yield higher agent satisfaction. The constraint of segregation to a node's six immediate connected neighbors implies that network structure influences the system predominantly through economic diffusion facilitated by commuting. To quantify this diffusive effect, we employ the second largest eigenvalue from the normalized laplacian matrix associated the the network $\lambda_2$, a metric that captures both network connectivity and relaxation dynamics \cite{Grabow(2010)}.

As we can see in the table \ref{TAB:landa2}, satisfaction increases with the rewiring probability $p$ for small-world networks, thus it allows for a greater diffusion of wealth among zones with different monetary levels. This diffusion is linked to the eigenvalue $\lambda_2$, and to the agents' own satisfaction, such that they grow together. In contrast, integration declines in response to a reduction in the number of neighboring nodes, affecting the segregation terms from eq. \ref{eq:Idis}. We proceed to explain this relationship across a range of $p$ values. When $p=0$, each node is exclusively connected to its six nearest neighbors, i.e. no long-range connections, blue agents cluster in the deprived zone, while red agents dominate the wealthier areas. Despite limited resources, blue agents reach almost $60\%$ satisfaction because red agents occasionally end up in poor zones and share their surplus wealth with nearby blues. This random placement boosts blue satisfaction, though the effect averages out across the system. For the purpose of segregation analysis, only these immediate neighbors are considered. As $p$ increases, local connections are diminished, while the number of long-range connections grows, leading to a decrease in overall integration. Let us take an intermediate value such as $p=0.4$. Each link has a $40\%$ chance of becoming a long-range connection, though nodes still retain part of their local neighborhood. As a result, blue agents spin (alignment) decreases, meaning more red agents appear in the blue zones and vice versa. The increase in long-range links also leads to more effective economic exchanges between rich and poor areas. For the case of random links, $p=1$, most nodes lose their local neighborhood and become isolated but still satisfied. Red agents are more dispersed, and satisfaction levels are now similar in both rich (red) and deprived (blue) zones. Finally, in a fully connected graph, each node has $6$ close neighbors and can exchange money with any other agent. This allows full communication across all zones. Blue agents remain in affordable areas, and both rich and poor zones are densely occupied, except at the borders.

\begin{table}[H]
\begin{center}
\begin{tabular}{|c|c|c|c|c|c|c|c|}
\hline
\textbf{}	& \textbf{$p=0$}	& \textbf{$p=0.2$}     & \textbf{$p=0.4$}  & \textbf{$p=0.6$}  & \textbf{$p=0.8$}  & \textbf{$p=1.0$} & Complete \\
\hline
\hline
\textbf{$\lambda_2$} & $0.04$ & $0.14\pm 0.03$ & $0.24\pm 0.03$ & $0.31\pm 0.03$ 
& $0.33\pm 0.02$ & $0.33\pm 0.02$ & $1.02$\\
\hline
\textbf{$\bar{H}$} & $0.66\pm 0.13$ & $0.70\pm 0.11$ & $0.70\pm 0.11$ & $0.69\pm 0.12$ 
& $0.71\pm 0.12$ & $0.78\pm 0.08$ & $0.86 \pm 0.07$\\
\hline
\textbf{$C_I$} & $0.22\pm 0.16$ & $0.18\pm 0.11$ & $0.15\pm 0.10$ & $0.11\pm 0.08$ 
& $0.06\pm 0.06$ & $0.007\pm 0.001$ & $0.23 \pm 0.12$\\
\hline
\end{tabular}
\end{center}
\caption{The probability of rewiring for small-world networks is given by $p$. Complete denotes the complete graph. The calculations are over the entire ensemble, considering $1000$ random generated networks for each measure. }
\label{TAB:landa2}
\end{table}

However, the complete graph configuration demonstrates optimality, with an observed eigenvalue consistent with the theoretical $(N+1)/N$ \cite{Das(2015)}. This network topology models a scenario of full inter-zonal connectivity within a city, thereby enhancing access to remunerative employment for economically constrained agents. The resulting direct diffusion between economic sources and sinks generates a high satisfaction index of $0.86 \pm 0.07$ and a comparatively elevated integration level of $0.23 \pm 0.12$. These values were derived using $q=0.005$, highlighting that the optimal urban configuration is characterized by a synergistic relationship between high inter-zonal connectivity and efficient, low-cost transportation.

\begin{figure}[H]
\center
\includegraphics[width=12 cm]{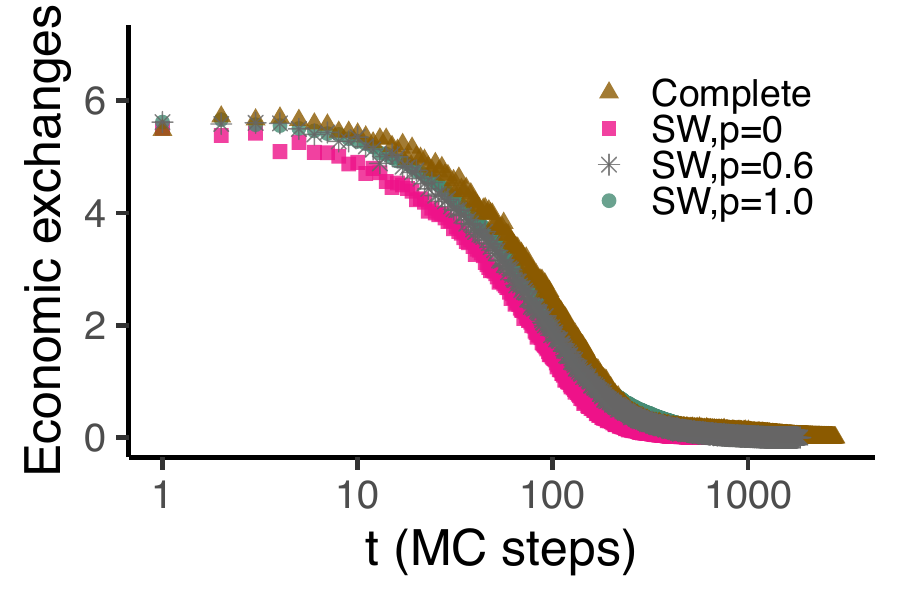}
\caption{Economic exchanges evolution for different networks, being $q=0.005$. Small-world networks with rewiring probabilty $p$ are written as "SW, p value". "Complete" denotes the complete graph.}
\label{Fig:evnet}
\end{figure}

We now turn our attention to the dynamic evolution of economic exchanges. As observed in Fig. \ref{Fig:evnet}, an increasing number of connections between affluent and economically deprived areas leads to a longer time for the system to reach equilibrium. These connections facilitate greater economic transfer between the two zones, enabling a more extensive redistribution of wealth. Consequently, the system takes longer to stabilize as the number of effective links between sources and sinks increase. Complete graph which maximizes the interconnectedness between different zones, represents the upper bound in terms of equilibration time.

\unskip

\section{Conclusions}
\label{sec:conclusion}

The extended Schelling model proposed in this study allows us to deepen the understanding of segregationist trends. Agents occupy nodes within a small-world network and can undertake two actions to increase their satisfaction: relocation to another spot and economic exchanges. We consider that $8\%$ of the positions within the network are vacant, and that there are two contrasting economic levels, high and low, for both the agents and the network nodes.

We found that although the initial agent distribution is random, they tend to form two distinct clusters. The economically advantaged group occupies the more expensive locations within the city, while those with financial constraints tend to occupy the cheaper locations, as illustrated in Fig. \ref{Fig:rueda1}. The model vacancies are located at the boundary between these two groups, further separating them and reflecting the real-world phenomenon where borders such as roads tend to segregate areas of different socioeconomic levels \cite{Mahajan(2024)}. Furthermore, wealthier agents typically exhibit higher happiness indices compared to those with financial needs. Transfer of economic exchanges occurs predominantly through connections between affluent areas and those in need of financial resources, discounting economic interactions between neighbors, who are often in similar economic circumstances as it is depicted in Fig. \ref{Fig:money}. This gives rise to a model that reproduces the spatial mismatch theory: individuals with economic needs must undertake long commutes to reach areas where they can obtain adequate wages, increasing the time taken to reach their workplace within the city.

Focusing on the model evolution, which can be observed in Fig. \ref{Fig:evol}, we have demonstrated that relocations follow power laws, while economic exchanges are governed by exponentials. Both can be explained by the different underlying processes: relocations reach myopic Nash equilibria \cite{Dallasta(2008), Vinkovic(2006)}, where agents typically do not find a better position after the initial moments, given the locations chosen by other agents. Following this initial phase, relocations tend to cease. Monetary exchanges are a diffusive process involving the exchange of economic units of size $q$. Wealthy agents act as sources, and individuals performing work function as sinks. These exchanges discontinue when the satisfaction level of either party decreases.

Having confirmed that our model reflected real-world scenarios, we focused on studying the impact of different parameters: the economic exchange size $q$ (\ref{subsec:qs}), the potential delay in agent actions (\ref{subsec:delays}), and the network topology (\ref{subsec:networks}). As it is depicted in Fig. \ref{Fig:qstate} \textbf{b)} low $q$ values imply low integration, but economic redistribution processes are effective, resulting in high overall happiness despite the high segregation. As $q$ increases, a transition occurred around $q=0.1$, and agents became integrated but unhappy. 

Regarding delays in actions, many models typically assume instantaneous agent relocation; however, in reality, any process associated with real estate markets or economic exchanges requires time. To account for this, we considered that half of the agents experienced a one-step delay in performing these processes. The delay effect on the satisfaction of wealthy agents was less significant than on those with economic disadvantages. In other words, agents with limited resources who experience delays in accessing nodes connected to wealthy areas will likely find it difficult to access economic resources that would improve their satisfaction (see Table \ref{Tab:delays}). When evaluating network types, we found that complete graphs, which connect all agents to each other, exhibited the highest level of mean satisfaction as it is observed in Table \ref{TAB:landa2}. This finding is consistent with the previous observations, as it allows for a greater wealth redistribution. Although not included in the previous results, we have also studied how the rise in system vacancies affects the system. It expands the edges between clusters, but it does not noticeably alter the other findings in this research.

In summary, this study builds upon a socio-physical model of segregation incorporating economic exchanges, which reproduces the spatial mismatch theory. After evaluating the various factors that can increase agent happiness, we found that it occurs in cities with high interconnectedness between affluent areas and those with economic needs (complete graph), and with cheap and efficient transportation, which we consider included within a small economic size $q$.

Future research should explore model variants incorporating city structures derived from Geographic Information Systems (GIS). Specifically, analyzing data from urban areas where segregation and commuting patterns are significant, such as Washington D.C. and New York City, would be valuable. 

In conclusion, we emphasize the critical importance of understanding the fundamental principles underlying segregation processes and how urban residents adapt to housing and labor markets, particularly in a world where pro-segregation attitudes are prevalent.

\section*{CRediT authorship contribution statement}
Diego Ortega: Conceptualization, Software, Visualization, Writing - original draft. Elka Korutcheva: Conceptualization, Supervision, Writing - review $\&$ editing.
\section*{Declaration of competing interest}
The authors declare that they have no known competing financial interests or personal relationships that could have appeared to influence the work reported in this paper.
\section*{Acknowledgments} 
We acknowledge financial support from the Spanish Government through grants PID2019-105182GB-I00 and PID2024-159024NB-C22. 
\section*{Declaration of generative AI and AI-assisted technologies in the manuscript preparation process}
During the preparation of this work the authors used Gemini in order to improve the readability. After using this tool, the authors reviewed and edited the content as needed and take full responsibility for the content of the published article.

\end{document}